# Photon counting spectroscopic CT with dynamic beam attenuator

Haluk Atak [1)] and Polad M. Shikhaliev [a)]
[1)]*Department of Nuclear Engineering, Hacettepe University, Ankara, Turkey*

## Abstract

**Purpose:** Photon counting (PC) computed tomography (CT) can provide material selective CT imaging at lowest patient dose but it suffers from suboptimal count rate. A dynamic beam attenuator (DBA) can help with count rate by modulating x-ray beam intensity such that the low attenuating areas of the patient receive lower exposure, and detector behind these areas is not overexposed. However, DBA may harden the beam and cause artifacts and errors. This work investigates positive and negative effects of using DBA in PCCT. **Methods:** A simple PCCT with single energy bin, spectroscopic PCCT with 2 and 5 energy bins, and conventional energy integrating CT with and without DBA were simulated and investigated using 120kVp tube voltage and 14mGy air dose. The DBAs were modeled as made from soft tissue (ST) equivalent material, iron (Fe), and holmium (Ho) K-edge material. A cylindrical CT phantom and chest phantom with iodine and $CaCO_3$ contrast elements were used. Image artifacts and quantification errors in general and material decomposed CT were determined. **Results:** Simple PCCT exhibited major image artifacts and quantification errors when DBAs were used. The artifacts and errors were decreased with 2bin spectroscopic PCCT and nearly eliminated with 5bin spectroscopic CT. The photon starvation noise did present with Fe-DBA due to strong absorption of lower energy photons. The 5bin PCCT with ST-DBA and Ho-DBA were nearly free of artifacts and photon starvation noise. The Ho-DBA better preserved low energy photons due to its K-edge at 55.6keV, which decreased beam hardening artifacts and improved material decomposition. The Ho-DBA was miniature having 1.4mm thickness and 2cm length, which is much smaller than ST-DBA and 10 times smaller than Fe-DBA. **Conclusion:** If successfully implemented, DBA fabricated from K-edge material such as Ho can address count rate problem of PCCT and provide miniature size, minimal image artifacts, and improved material decomposition at lowest patient dose.

[a)] Corresponding author: pshikhal@yahoo.com

## 1. Introduction

Photon counting computed tomography (PCCT) can provide lowest patient dose and material selective CT imaging predicted at early stages of CT technology (Hounsfield 1973). However, despite intensive investigations (Shikhaliev et al. 2005; Schlomka et al. 2008; Shikhaliev 2008; Taguchi et al. 2010; Le and Molloi 2011; Shikhaliev and Fritz 2011; Silkwood et al. 2013; Atak and Shikhaliev 2015; Yu et al. 2016), PCCT is still suboptimal for clinical CT imaging. One major problem is the limited count rate of the x-ray detectors used in PCCT systems. In PCCT, each x-ray photon should be counted separately at very high rates, and its energy should be measured.

Several approaches exist to address the count rate problem in PCCT. First, the detector material could be advanced to improve its charge generation, transport, collection, and defect characteristics. However, these methods require conceptual breakthroughs. Second, the detector pixels could be divided into smaller sub-pixels to decrease the pixel count rate at the same x-ray flux to the detector surface (Kappler et al. 2014; Yu et al. 2016). However, detectors with smaller sub-pixels suffer from charge sharing effects (d'Aillon et al. 2006; Kuvvetli and Jørgensen 2007; Shikhaliev et al. 2009; Faby et al. 2016).

Another approach is spatial modulation of the x-ray beam intensity so that the less attenuating parts of the object receive lower x-ray exposure. In this way, the x-ray flux along the detector surface could be minimal and approximately constant. This method was used in early CT systems when the object was placed in a water tank to flatten the x-ray intensity at the detector plane (Hounsfield 1973). Its simplified version is used in current CT systems as "bow-tie" filters that partially flattens the x-ray intensity at the detector surface (Webb 1988; Bushberg et al. 2002). Notice also that spatial modulation of x-ray intensity can optimize patient dose, address detector dynamic range problem, and decrease detected scatter which could further improve signal to noise ratio. This method was previously investigated also for projection x-ray imaging (Hasegawa et al. 1986; Vlasbloem and Kool 1988; Xu et al. 2004).

The spatial modulation of the beam intensity was recently used in PCCT where full flattening of the intensity at the detector surface was provided (Shikhaliev 2008; Shikhaliev and Fritz 2011; Shikhaliev 2012; Silkwood et al. 2013; Shikhaliev 2015). The above modulators had predetermined shapes designed for imaging round object. The modulator design reported in (Shikhaliev 2012; Silkwood et al. 2013; Shikhaliev 2015) can be used for clinical PCCT such as dedicated breast CT where breast is imaged in pendant geometry and can be shaped to a round shape. However, in whole body CT the attenuation pattern of the human body is complex and depends on projection angle. Therefore, a dynamically re-shapeable beam attenuator is need that could change its shape during CT gantry rotation. Such a dynamic beam attenuator (DBA) was proposed and investigated in (Szczykutowicz and Mistretta 2012; Szczykutowicz and Mistretta 2013; Szczykutowicz and Mistretta 2013; Szczykutowicz and Mistretta 2014). It includes a series of iron (Fe) wedges with rectangular cross-sections installed at the x-ray tube output. The wedges move back and forth during the CT scan and provide dynamically changing stepwise attenuation profiles. Another group reported similar DBA made from Fe except they used the wedges with triangular cross sections that could provide smooth attenuation profile with the intention that it could avoid discontinuities in the attenuation profiles associated with rectangular wedges used by previous group that could cause imaging artifacts (Hsieh and Pelc 2013; Hsieh et al. 2014).

The DBA fabricated from high-Z material such as Fe would predominantly attenuate the low energy photons and harden the beam. Furthermore, the beam hardening could be spatially non-uniform because less attenuating parts of the object require more Fe filtration, and vice-versa. Therefore, beam hardening artifacts may appear in CT images, and CT number and material decomposition errors may also present. Recently, it has been proposed and experimentally proven that using K-edge filtered x-ray beams in PCCT can substantially improve material decomposition (Shikhaliev 2012). The filter material with K-edge energy located around the middle of the x-ray spectrum provides two major advantages: (1) Separating the x-ray spectrum in two parts which improves material decomposition and (2) decreasing attenuation of the low energy part of the spectrum which minimizes the beam hardening artifacts and decomposition errors. The above work (Shikhaliev 2012) used the K-edge filters with uniform thicknesses attached to an acrylic beam intensity modulator (i.e., the static version of DBA). The combination of the K-edge filter and intensity modulator together improved spectral separation and addressed the count rate problem of PCCT. However, the intensity modulator could also be fabricated from a K-edge material or its mixture to address simultaneously spectral separation and beam intensity modulation. The work (Hsieh and Pelc 2015) theoretically investigated PCCT with K-edge DBA and has shown that K-edge DBA improves count rate performance and noise variance as compared to PCCT with conventional bowtie filter.

The purpose of the current study was to perform further investigations of the key features of PCCT with DBA that have not been investigated previously. These include: **(1)** Investigations of the beam hardening artifacts and associated quantification errors in PCCT with DBAs made of different materials including high-Z material (Fe), K-edge material (Ho), tissue-equivalent material (soft tissue), and with no DBA; **(2)** Investigations and quantifications of the image artifacts associated with stepwise and smooth attenuation profiles of the DBAs; **(3)** Quantitative comparisons of contrast to noise ratios (CNR) of clinically relevant contrast agents iodine and calcifications with different DBAs; **(4)** Inter-comparisons between PCCT and energy integrating CT with different DBAs; **(5)** Investigations of the image artifacts with different DBAs using non-spectroscopic (simple), sub-spectroscopic (2bin) and fully spectroscopic (5bin) PCCT systems; **(6)** Investigations of the effect of tube focal spot size on PCCT image artifacts with different DBAs. The conventional (not decomposed) PCCT images with DBAs were also investigated in addition to material decomposed images to assure that the quality of the conventional PCCT images do not deteriorate when different DBAs are used.

## 2. Methods and Materials

### 2.1. PCCT system with DBA

#### 2.1.1. System configuration

To simulate PCCT with DBA, the system configuration of the Siemens Somatom Definition Flash CT (NHS 2009) was used because a prototype PCCT based on this system has recently been developed (Yu et al. 2016). This system used CdTe photon counting (PC) detector, and the count rate problem was improved using smaller detector pixels with 0.25x0.25mm$^2$ sizes (Yu et al. 2016). However, in our study we used conventional pixel size instead of small pixels, and the count rate problem was addressed using DBA. The PCCT with DBA had a source to isocenter distance of 59.5cm, a detector to isocenter distance of 48cm, and a magnification factor of 1.82 **(figure 1)**.

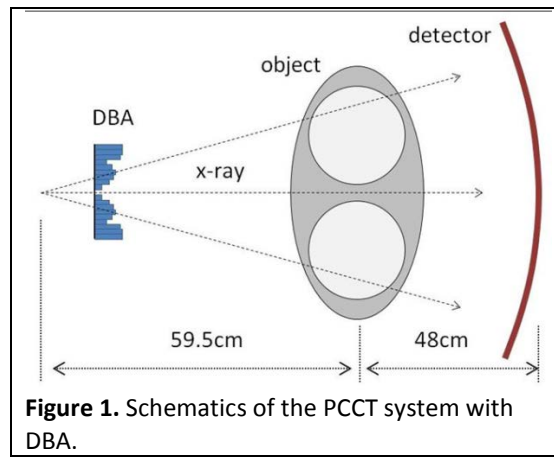

**Figure 1.** Schematics of the PCCT system with DBA.

#### 2.1.2. Photon counting detector

The system included energy selective PC detector with 760 pixels per detector row. The pixel size was 1.1mm along Z-axis and 1.4mm along the row. The 1.82 magnification provided 0.77mm and 0.6mm pixel sizes in XY plane and Z direction, respectively, measured at the isocenter.

Four types of CT detectors were used: (1)Simple PC detector with single energy threshold which can count the x-ray photons without energy information (Shikhaliev et al. 2004; Shikhaliev et al. 2005; Aslund et al. 2006; Aslund et al. 2007); (2)PC detector with two energy thresholds that can split the photons into two energy bins and allow two-material decomposition (Yu et al. 2016);

(3)PC detector with 5 energy bins that can provide multi-energy CT acquisition and multi-material decomposition (Schlomka et al. 2008; Shikhaliev 2008; Shikhaliev and Fritz 2011; Shikhaliev 2012; Atak and Shikhaliev 2015; Shikhaliev 2015); and (4)Conventional energy integrating detector. All detectors were considered to be ideal. The real PC and energy integrating detectors have numerous limitations. These limitations have complex physical mechanisms and are subject for separate studies (d'Aillon et al. 2006; Kuvvetli and Jørgensen 2007; Heismann et al. 2008; Shikhaliev et al. 2009; Faby et al. 2016). It is particularly difficult to correctly model and simulate real PC detectors because they suffer from charge sharing and pixel crosstalk, K-x-ray escape, double and multiple counting of the same x-ray photon, and spectral distortion due to the carrier trapping.

### 2.1.3. Count rate requirements

The count rate requirements and pixel count statistics for the whole body PCCT system can be estimated assuming $1.1 \times 1.4 mm^2$ detector pixel size, 120kVp tube voltage, 0.35mm Cu filtration, 200mA tube current, 30cm patient thickness, 0.5s gantry rotation time, and number of CT projection 720. Based on these parameters, the unattenuated x-ray flux at the detector surface is approximately $3 \times 10^8$ photon/mm$^2$/s (Poludniowski 2007; Poludniowski and Evans 2007) that corresponds to 450Mcount/pixel/sec count rate, which is too high with existing PC detectors. However, the beam passing 30cm soft tissue will be attenuated by approximately 500 times and the count rate will be approximately 1Mcount/pixel/sec. The ideal DBA is designed such that it does not attenuate the beam passing the thickest part of the object, and attenuates the beam over the thinner parts so that the count rate along the detector surface is 1Mcount/pixel/sec. However, the DBA decreases the average x-ray exposure to the object. This decrease can be compensated by increasing the tube current so that the average exposure remains same as without DBA. As will be shown in the Results section, this compensation will require increasing the tube current by approximately 3 times. Therefore, the required detector count rate with an ideal DBA is 3Mcount/pixel/sec. The previously used CdZnTe and CdTe detectors provided up to 2Mcount/pixel/sec count rates (Schlomka et al. 2008; Shikhaliev 2008; Shikhaliev and Fritz 2011; Shikhaliev 2012; Shikhaliev 2015). However, current detectors can provide up to 5Mcount/pixel/sec count rates (Taguchi and Iwanczyk 2013). Therefore, current PC detectors can satisfy the count rate requirements for clinical whole body PCCT systems with DBA.

### 2.2. Phantoms

A cylindrical soft tissue phantom with 30cm diameter was used for quantitative evaluation of the PCCT with DBA. The phantom included iodine (10mg/cm$^3$), CaCO$_3$ (150mg/cm$^3$), adipose, and water contrasts (**figure 2a**).

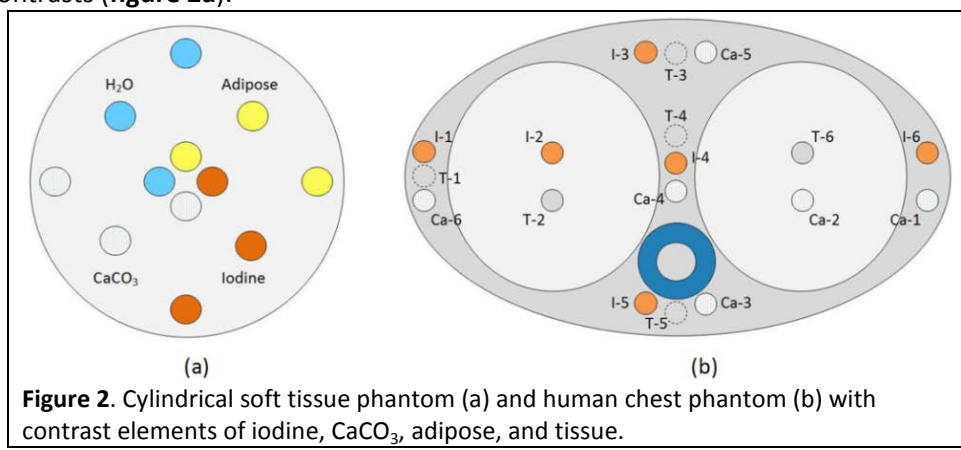

**Figure 2**. Cylindrical soft tissue phantom (a) and human chest phantom (b) with contrast elements of iodine, CaCO$_3$, adipose, and tissue.

The contrast elements had a cylindrical shape and 3cm diameter. They were located at 3cm, 9cm, and 12.5cm distances from the center of the phantom to quantify signal uniformities. The second phantom was a human chest phantom with 38cm length and 23cm height (**figure2b**). It included simulated lungs, mediastinum, and contrast elements. The lung material was the same as soft tissue but had 20% density of the soft tissue. The iodine (I-1 – I-6), $CaCO_3$ (Ca-1 – Ca-6), and soft tissue contrasts (T-1 – T-6) were distributed at different locations to quantify signal uniformities.

### 2.3. Design and simulations of DBA

#### 2.3.1. Material selection for the DBA

An ideal DBA should be fabricated from tissue equivalent material to minimize the beam hardening effects. Also, DBA should be compact and compatible with output port of the x-ray tube. However, soft tissue equivalent DBA would be impractical because its thickness will be comparable to the patient thickness. The Fe-DBA investigated in previous studies had a maximum thickness of 1.5cm (Hsieh and Pelc 2013; Szczykutowicz and Mistretta 2013; Szczykutowicz and Mistretta 2013; Hsieh et al. 2014; Szczykutowicz and Mistretta 2014; Hsieh and Pelc 2015). However, Fe strongly hardens the x-ray beam, which may result in beam hardening artifacts and material decomposition errors. The K-edge filter material recently used in experimental PCCT (Shikhaliev 2012) separates the x-ray spectrum in two parts improving material decomposition, and minimally absorbs the low energy photons decreasing beam hardening artifacts and decomposition errors. In the current work, DBAs were modeled as made from Fe, Ho K-edge material (atomic number 67 and K-edge energy 55.6keV), and soft tissue equivalent material (ICRU-44), and used for spectroscopic PCCT simulations.

**Figure 3** shows the energy spectra of the x-ray beams filtered by three DBA materials: soft tissue with 0, 5, 15, and 30 cm thicknesses, Fe with 0, 0.95, 4.9, and 14mm thicknesses, and Ho with 0, 0.18, 0.66, and 1.55mm thicknesses. The thicknesses of ST, Fe, and Ho filters were chosen such that corresponding thicknesses provide same attenuation of the air dose (kerma).

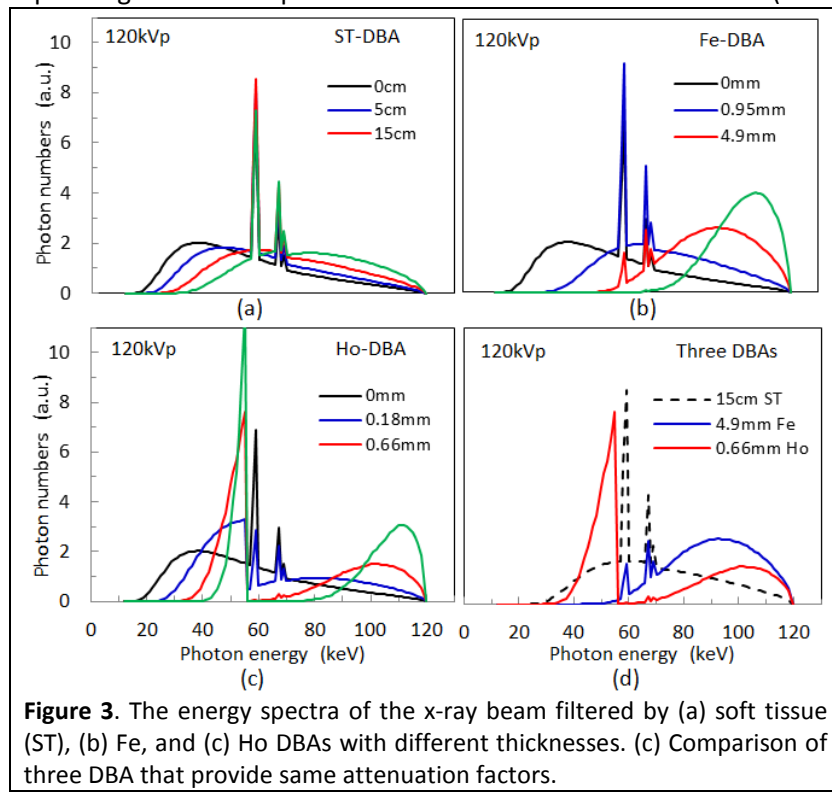

**Figure 3**. The energy spectra of the x-ray beam filtered by (a) soft tissue (ST), (b) Fe, and (c) Ho DBAs with different thicknesses. (c) Comparison of three DBA that provide same attenuation factors.

### 2.3.2. Simulation of DBA

The DBA is designed as 1D array of individual filter elements (wedges) installed at the output port of the x-ray tube (**figure 1**). Each wedge has a thickness changing from zero to maximum (Hsieh and Pelc 2013; Szczykutowicz and Mistretta 2013). When CT gantry is rotated and CT projections are acquired, the wedges are moved in the direction perpendicular to the CT gantry plane, and DBA dynamically changes its attenuation profile. This provides matching the shape of the DBA to the attenuation pattern of the object at different projection angles. For an ideal DBA, the intensity of the x-ray beam passed through DBA and object should be constant along the detector rows at all projection angles. To calculate DBA thickness distributions at each projection angle, the attenuation patterns (line integrals) of the object should be determined. This data is generated using a preliminary low-dose CT scan before the main CT acquisition is performed.

To calculate the DBA thickness distribution, the mu-maps of the phantoms were forward projected to generate the line integrals, assuming 720 CT projections in 360 degree rotation angle, and 760 rays (a ray per detector pixel). Among all rays at all projection angles, the ray passing through the maximum attenuating part of the object was identified, the photon number $N_0$ passed along this ray was calculated, and zero thickness of the DBA was assigned to this particular ray. Then the DBA thicknesses were calculated for other rays at all projection angles such that the photon numbers passed the DBA and phantom along all rays equal to $N_0$. The wedge thickness $t_K(\alpha, \theta)$ corresponding to the ray at the fan angle $\theta$ and projection angle $\alpha$ was determined by numerical solution of the equation

$$\int_{E_{\min}}^{E_{\max}} N(E) e^{-\int \mu(E, \alpha, \theta, r) dr - t_K(\alpha, \theta) \mu_K(E)} dE = N_0 \qquad (1)$$

where $N(E)$ is the photon distribution in the original x-ray beam, $E_{\min}$ and $E_{\max}$ are the lowest and highest photon energies in the beam, $\mu(E, \alpha, \theta, r)$ is the linear attenuation coefficient (LAC) of the object at energy $E$, and along the ray $r$ at the fan angle $\theta$ and projection angle $\alpha$, and $\mu_K(E)$ is the LAC of the DBA material at energy $E$. The minimal number of photons $N_0$ is determined as

$$N_0 = \int_{E_{\min}}^{E_{\max}} N(E) e^{-\int \mu(E, \alpha_0, \theta_0, r_0) dr_0} dE \qquad (2)$$

where $r_0$ is the unique ray at fan angle $\theta_0$ and projection angle $\alpha_0$ that exhibits largest photon attenuation.

The simulations were performed for an ideal DBA with 760 wedges (a wedge per ray), and for a realistic DBA with 38 wedges. To calculate 38-wedge DBA, the original projection data of the object was down sampled from 760 to 38 steps within the same fan angle. Because fabrication of DBAs with large numbers of wedges is challenging, tradeoff is allowed. For example, the works (Szczykutowicz and Mistretta 2013; Hsieh et al. 2014) used 15 wedges made of Fe covering similar fan angle. The DBAs with 760 and 38 steps were called *smooth* and *stepwise* DBAs, respectively.

Assuming 720 CT projection with 0.5 degree angular steps, and 0.5sec gantry rotation time, the DBA should reshape 1480 times per second. However, in practice, it is sufficient to compensate the slow changes in attenuation patterns associated with position and shape of the lungs, large bones, and peripheries of the body. Therefore, the shape of the DBA may change with larger angular steps. We simulated smooth and stepwise changing of the shape of DBA during the gantry rotation. To simulate stepwise change, the data was down sampled in the projection angle domain from 720 to 36 providing changes of the DBA shape 36 times in 360 degree rotation angle.

In the real CT system, the focal spot burring may affect performance of the DBA which is installed at the close distance from the focal spot. To test this effect, the PCCT images were also simulated with 1mm focal spot size, using stepwise DBA installed at 10cm from the focal spot and system geometry shown in figure 1. With the above parameters, the focal spot blurring of the DBA is 8 pixels with 0.77mm pixel size measured at the isocenter, and its effect on image artifacts was evaluated.

## 2.4. Simulation of PCCT with DBA

The 120kVp x-ray spectrum filtered with 3mm Al was generated using the method described in (Poludniowski 2007; Poludniowski and Evans 2007). The spectrum was divided into sub-regions with 3keV width, and the average photon energy $E_i$ was calculated for each sub-region. The distribution of the LACs within the phantom (mu-maps) was generated at each energy $E_i$ using known LACs (Hubbell and Seltzer 1995). For each energy $E_i$, the mu-map of the phantom was forward projected with 0.5° angular increments to create 720 CT projections in 360 rotation angle. Each CT projection included 760 line integrals corresponding to the number of rays. Using these line integrals, the x-ray attenuation factors along each ray were determined. The numbers of the x-ray photons arrived at the detector pixel were calculated in the absence of the phantom using 14mGy air kerma (Johns and Cunningham 1983). Then, the photon numbers per pixel were determined in the presence of the phantom using calculated attenuation factors for each ray. Finally, the count distribution per pixel was randomized assuming Poisson distribution of the detected photons. This procedure was repeated for each of the energy $E_i$. Thus, the quasi-monoenergetic CT projections with statistical count distributions were generated at 33 energies in the energy range of 20 - 120keV.

The above monoenergetic projection data was used to reconstruct CT images. To generate simple PCCT projection data, the 33 monoenergetic projections were summed providing polyenergetic CT projections in 20-120keV range. To generate 2bin PCCT projections, the monoenergetic projections were summed for 20-58keV and 59-120keV ranges. To create 5bin PCCT projections, the monoenergetic projections were summed in the 5 energy ranges (bins) 20-40keV, 40-58keV, 58-80keV, 80-100keV, and 100-120keV. The energy integrating CT data was generated similarly to simple PCCT with the difference that before summing the photon counts at each energy $E_i$ the counts were multiplied by corresponding energy $E_i$. The CT images were reconstructed from the above projection data using filtered backprojection method.

In the case when the DBA is installed at the x-ray tube output, the line integrals of the DBA were calculated similarly to that for the phantom. The CT acquisition with DBA was simulated to be as it would occur in real CT acquisition. First, the CT projections were simulated with DBA and object in the beam. Then, the object was removed, and CT projections were acquired with DBA in the beam. Then, using these two dataset the line integrals of the object alone were calculated and used for CT reconstruction. Notice that statistical noise was added to only CT projections acquired with DBA and object in the beam, and no statistical noise was added when CT projections were acquired with DBA alone in the beam. This is a reasonable assumption because the data with DBA alone can be acquired with low statistical noise. The low noise in DBA alone data can be achieved by (1)averaging the DBA alone data over multiple detector rows, (2)acquiring multiple consecutive scans with DBA alone and averaging, and (3)filtering DBA alone data to decrease high frequency statistical noise.

The PC and energy integrating CT images without DBA were also reconstructed using the same air dose as with DBA, measured at the isocenter. The flowchart of the simulation study is shown in **figure 4**.

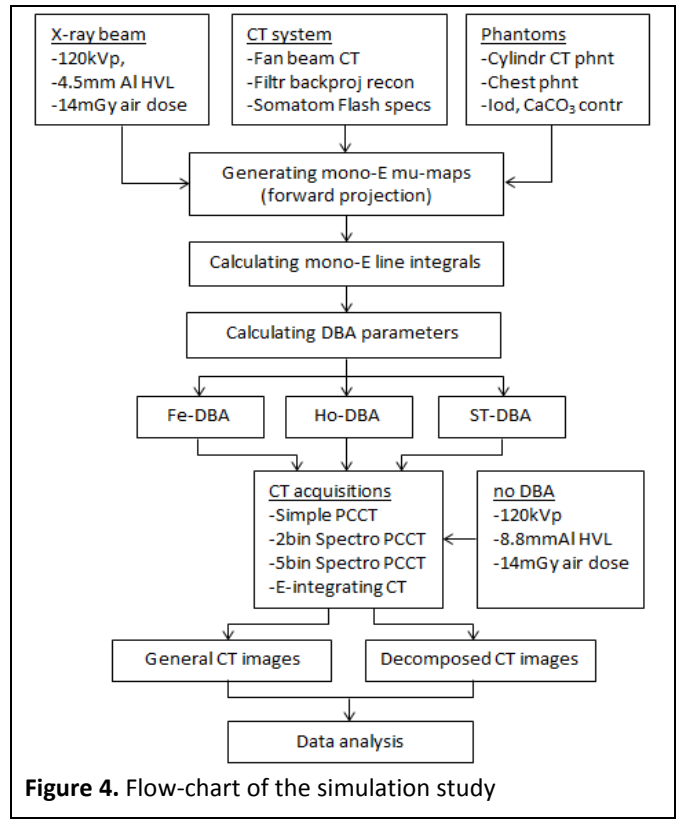

**Figure 4.** Flow-chart of the simulation study

## 2.5. Data analysis

While the reconstructed general CT images were immediately available for analysis, the 2bin and 5bin spectroscopic PCCT images were generated by optimal image domain weighting of the corresponding images for bin1-bin5 (Shikhaliev 2010).

The material decomposed (soft tissue background cancelled) images were generated using dual energy subtraction. For 2bin PCCT, the low and high energy bins 20-58keV and 58-120keV, respectively, were used. For 5bin PCCT, the low energy bin was composed by optimal weighting of the bin1 (20-40keV) and bin2 (40-58keV), and the high energy bin was composed by optimal weighting of the bin3 (58-80keV), bin4 (80-100keV) and bin5 (100-120keV) images.

For cylindrical CT phantom, dependences of the beam hardening artifacts on DBA material were determined quantitatively by measuring the magnitudes of the artifacts. The relative image noise was measured at different locations in the phantoms to quantify noise non-uniformity. The relative noise was determined as standard deviation of the pixel value divided by mean signal within the circular regions of interests (ROI) with 2cm diameters. For LAC measurements the images were low-pass filtered to suppress the statistical noise, and then the pixel values were averaged over 300 pixels within the ROIs. The resulted statistical errors were negligible as compared to measured LAC and non-uniformities associated with artifacts. The statistical noise was measured after low frequency components of the images were removed by subtracting low-pass filtered versions of the original images. Therefore, effects of statistical noise and low frequency background on measurements of the LAC and noise, respectively, were negligible.

Because step-wise DBA is more feasible way for realizing DBA concept, it was important to compare CT images acquired with step-wise DBA and smooth DBA to assess whether additional artifacts appear when step-wise DBA is used.

The qualitative comparisons were performed using color and black-white images, as well as by plotting image profiles that were particularly useful for visualizing and comparing the beam hardening artifacts. The true LACs were also calculated for the chest phantom for comparison using 120kVp tube voltage filtered by 3.6mm Fe and 10cm soft tissue for Fe-DBA case, and 0.43mm Ho and 10cm soft tissue for the Ho-DBA case, respectively. The above thicknesses of the Fe and Ho filters represent average thicknesses of the Fe-DBA and Ho-DBA, respectively, used in PCCT for imaging chest phantom.

## 3. Results

**Table 1** shows the key parameters of the DBAs simulated for cylindrical CT phantom. The maximum thicknesses of the Fe-DBA, Ho-DBA, and ST-DBA were 14.4mm, 1.55mm, and 300mm, respectively. It is noticeable that the maximum thickness of the Ho-DBA was much smaller than Fe-DBA. Because the length of the DBA wedges is approximately 10 times of its maximum thickness (Hsieh and Pelc 2013; Szczykutowicz and Mistretta 2013; Szczykutowicz and Mistretta 2013; Hsieh et al. 2014; Szczykutowicz and Mistretta 2014), the Ho-DBA could have a length of approximately 2cm. Therefore, a miniature DBA can be fabricated from Ho or other K-edge materials. The maximum HVL of the x-ray beams passed through the thickest parts of the DBA were 15.2mm, 11mm, and 10.9mm Al equivalent, for Fe-DBA, Ho-DBA, and ST-DBA, respectively. The mean HVL calculated by averaging over the fan angle were 10.3mm, 7.1mm, and 6.8mm Al equivalent for Fe-DBA, Ho-DBA, and ST-DBA, respectively. Thus, the average HVL of the x-ray beams after DBAs were comparable to HVL used in clinical CT systems (NHS 2009). The attenuation of the air dose was calculated for each wedge of the DBA and averaged over the fan angle, resulting in average beam attenuation factors of 3.33, 2.59, and 2.50, for Fe-DBA, Ho-DBA, and ST-DBA, respectively.

**Table 1.** DBA parameters for cylindrical CT phantom

| DBA | Fe | Ho | ST |
|---|---|---|---|
| $t_{max}$ (mm) | 14.4 | 1.55 | 300 |
| <t> (mm) | 2.65 | 0.29 | 67 |
| $HVL_{max}$ (mm Al) | 15.2 | 11.0 | 10.9 |
| <HVL> (mm Al) | 10.3 | 7.1 | 6.8 |
| <Beam atten.> | 3.33 | 2.59 | 2.50 |

**Figure 5** shows the sinogram representation of CT projections including photon count distributions with Fe-DBA and without DBA. When no DBA is used, the photon counts inside and outside of the object may differ by hundreds of times, while with DBA the average counts inside and outside of the object are similar.

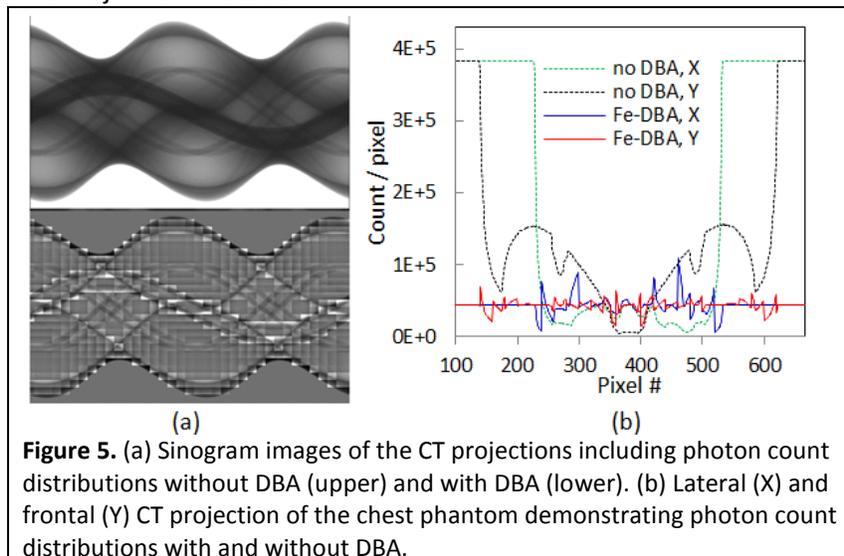

**Figure 5.** (a) Sinogram images of the CT projections including photon count distributions without DBA (upper) and with DBA (lower). (b) Lateral (X) and frontal (Y) CT projection of the chest phantom demonstrating photon count distributions with and without DBA.

**Figure 6** shows the images of the empty CT phantom acquired with simple PCCT, 2bin PCCT, and 5bin PCCT, and using Fe-DBA, Ho-DBA, and ST-DBA.

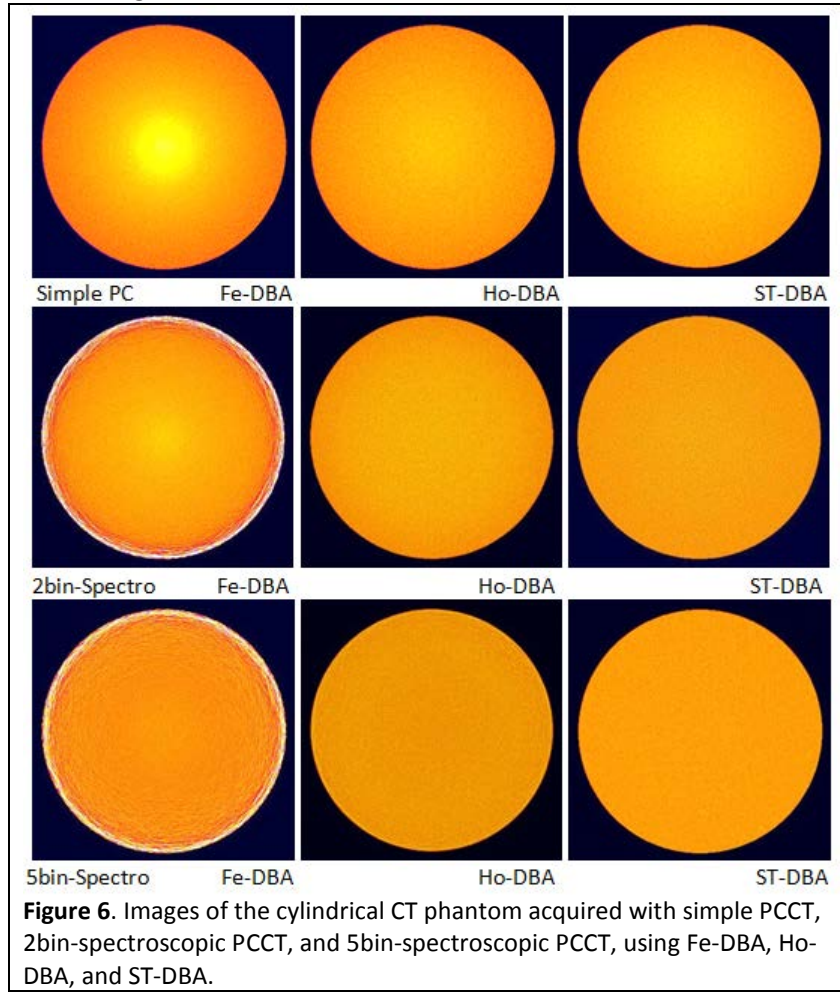

**Figure 6**. Images of the cylindrical CT phantom acquired with simple PCCT, 2bin-spectroscopic PCCT, and 5bin-spectroscopic PCCT, using Fe-DBA, Ho-DBA, and ST-DBA.

**Figure 7** shows corresponding image profiles of the above CT images. As can be seen, simple PCCT exhibits largest beam hardening artifacts with convex ("roof") like shape as opposed to well-known beam hardening "cup" artifacts without DBA. The artifacts are decreased with 2bin PCCT, and nearly eliminated with 5bin PCCT. However, with 5bin PCCT and Fe-DBA the noise is elevated toward the periphery of the phantom because increased thickness of Fe-DBA at larger fan angles resulted in photon starvation in lower energy bins (see **figure 2b**). The Ho-DBA is free of this limitation, and it performs as good as ST-DBA because Ho K-edge material preserves the low energy parts of the beam below the K-edge energy of 55.6keV, and photon starvation effect at lower energies is not substantial (**figure 2c**).

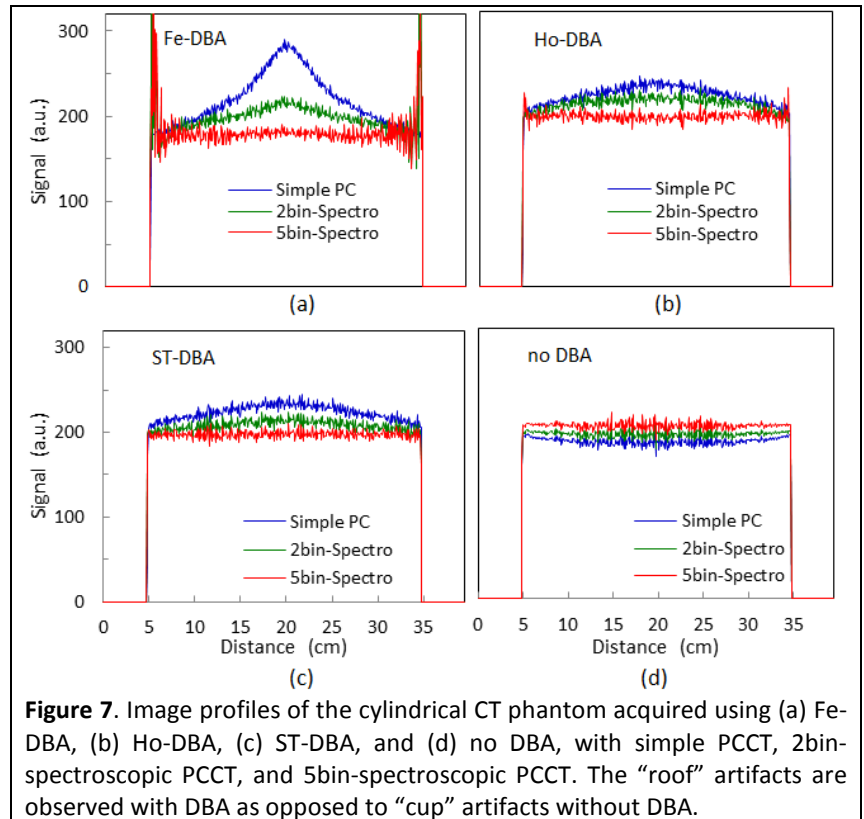

**Figure 7**. Image profiles of the cylindrical CT phantom acquired using (a) Fe-DBA, (b) Ho-DBA, (c) ST-DBA, and (d) no DBA, with simple PCCT, 2bin-spectroscopic PCCT, and 5bin-spectroscopic PCCT. The "roof" artifacts are observed with DBA as opposed to "cup" artifacts without DBA.

**Table 2** includes the magnitudes of "cupping" and "roofing" artifacts. The artifacts in simple PCCT images with Fe-DBA are by a factor of 3-4 larger than that with Ho-DBA and ST-DBA. Thus, Ho-DBA has a major advantage over Fe-DBA. The artifacts are decreased by a factor of 2 for all DBAs when 2bin spectroscopic PCCT is used while Fe-DBA still exhibits largest artifacts. Finally, in 5bin spectroscopic PCCT, the artifacts are decreased by a factor of 10-20 compared to simple PCCT, and can be neglected.

**Table 2.** Magnitudes of the cupping artifacts in CT phantom images

| Cupping magnitude | Fe-DBA | Ho-DBA | ST-DBA | no DBA |
|---|---|---|---|---|
| Simple PC | ±20% | ±6.2% | ±5.0% | ±1.5% |
| 2bin-Spectro | ±10% | ±4.6% | ±3.0% | ±0.6% |
| 5bin-Spectro | ±1.7% | ±0.3% | ±0.3% | ±0.1% |

The relative image noises at 3cm, 9cm, and 12.5cm distances from the phantom center are presented in the **Table 3**. For simple PCCT, the average noises are similar for Fe-DBA, Ho-DBA, and ST-DBA, as well as for the images acquired without DBA because all four CT acquisitions used same air dose. However, for 2bin and 5bin spectroscopic PCCT with Fe-DBA the noise is increased by a factor of 2 and 10, respectively, at the periphery of the phantom due to the photon starvation effect. The noise non-uniformity in CT images acquired without DBA is well-known as more photons pass the periphery of the phantom providing less statistical noise as compared to central parts.

**Table 3.** Relative noise in cylindrical phantom image

| Relative noise % 3/9/12.5cm | Fe | Ho | ST | No DBA |
|---|---|---|---|---|
| Simple PC | ±5.1/4.8/ 3.9 | ±5.4/5.6/5.3 | ±5.5/5.8/5.2 | ±7.0/6.1/3.9 |
| 2bin-Spectro | ±4.7/5.0/11 | ±5.5/5.6/5.2 | ±6.1/6.4/5.6 | ±7.2/6.3/4.0 |
| 5bin-Spectro | ±7.9/14/47 | ±6.4/6.0/6.8 | ±6.2/6.3/5.3 | ±7.3/6.4/4.1 |

**Figure 8** shows effect of the step-wise DBA design on image quality when 38 wedges (a wedge per 20 consecutive detector pixels) were used instead of 760 wedges (a wedge per pixel).

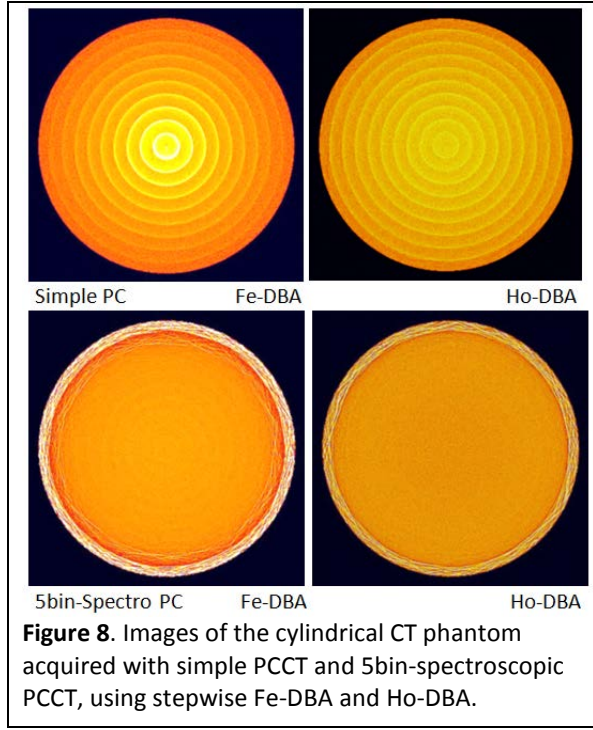

**Figure 8.** Images of the cylindrical CT phantom acquired with simple PCCT and 5bin-spectroscopic PCCT, using stepwise Fe-DBA and Ho-DBA.

The ring artifacts appear at the top of the "roof" artifacts (**figure 9**). The ring spikes appear due to the stepwise changes of the DBA thicknesses. As can be seen from figure 9, using non-zero (1mm) focal spot size decreases the ring artifacts with Fe-DBA and Ho-DBA because sharp edges of the DBA are smoothed due to focal spot blurring. The magnitudes of the ring artifacts with stepwise Fe-DBA are approximately 30% and 7% with zero and 1mm focal spot, respectively. Corresponding values for stepwise Ho-DBA are 15% and 5%, respectively. In 5bin spectroscopic PCCT, stepwise Fe-DBA and Ho-DBA provide negligible ring artifacts with magnitudes comparable to statistical noise.

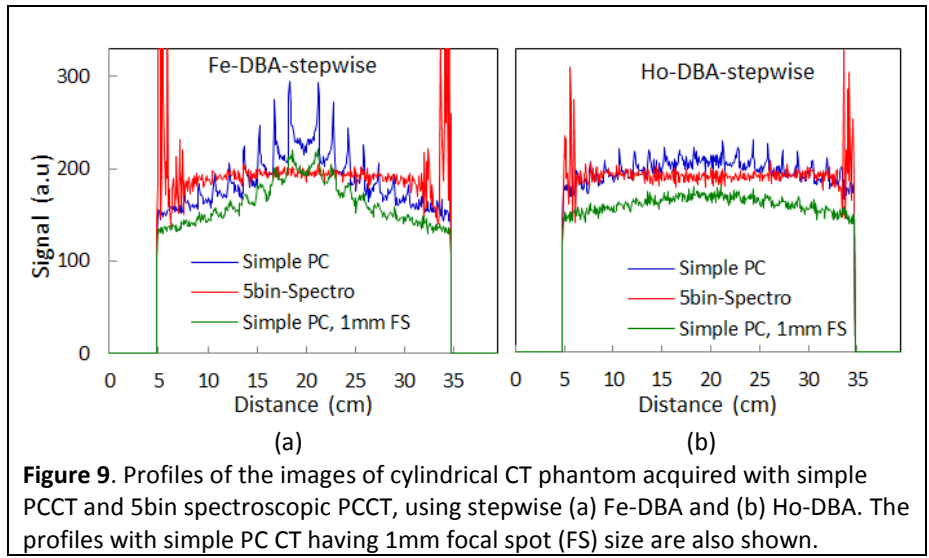

**Figure 9**. Profiles of the images of cylindrical CT phantom acquired with simple PCCT and 5bin spectroscopic PCCT, using stepwise (a) Fe-DBA and (b) Ho-DBA. The profiles with simple PC CT having 1mm focal spot (FS) size are also shown.

**Figure 10** shows material decomposed (soft tissue cancelled) images of the contrast elements. The beam hardening artifacts in low and high energy images propagate to material decomposed images after dual energy subtraction. The magnitudes of these artifacts correlate with magnitudes of "roof" artifacts given in the **Table 2**.

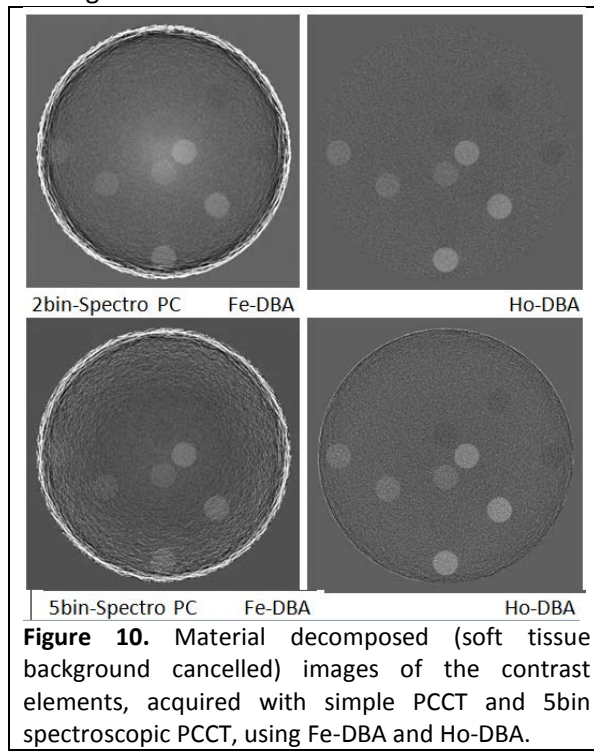

**Figure 10.** Material decomposed (soft tissue background cancelled) images of the contrast elements, acquired with simple PCCT and 5bin spectroscopic PCCT, using Fe-DBA and Ho-DBA.

**Figure 11** shows background profiles of the tissue cancelled images. The background signal should be zero, but it substantially deviates from zero, and is non-uniform with elevated noise when Fe-DBA is used. These limitations are nearly eliminated when Ho-DBA and ST-DBA are used. When no DBA is used signal amplitude is uniformly zero, but the noise is non-uniform due to the stronger photon absorption over the central parts of the phantom.

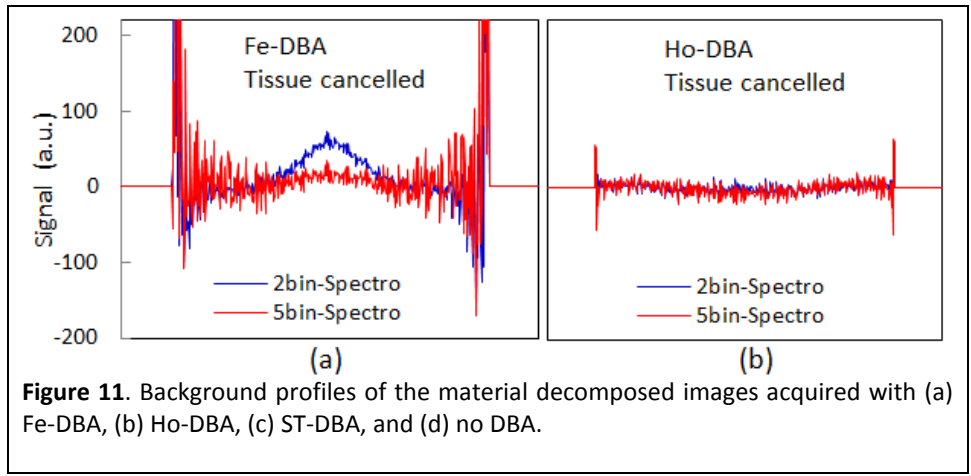

**Figure 11**. Background profiles of the material decomposed images acquired with (a) Fe-DBA, (b) Ho-DBA, (c) ST-DBA, and (d) no DBA.

Because both contrast and noise are affected by DBA, the CNR values of the contrast elements in decomposed images were calculated and presented in **Table 4**. The CNR are non-uniform with Fe-DBA and with no DBA, and Ho-DBA and ST-DBA provide better CNR uniformity.

**Table 4.** CNR at different positions in tissue-cancelled images

| CNR at 3.0/9.0/12.5cm | Fe | Ho | ST | no DBA |
|---|---|---|---|---|
| 2bin-Spectro -Iod | 2.8/2.3/1.1 | 2.5/2.9/3.2 | 2.3/2.2/1.9 | 1.4/2.6/3.3 |
| 2bin-Spectro -CaCO$_3$ | 1.4/0.9/0.5 | 1.1/1.3/1.4 | 1.0/1.0/1.0 | 0.8/1.1/1.4 |
| 5bin-Spectro -Iod | 2.7/1.5/0.6 | 2.6/2.8/3.0 | 2.4/2.5/2.7 | 1.8/2.5/3.3 |
| 5bin-Spectro -CaCO$_3$ | 1.2/0.6/0.2 | 1.0/1.3/1.3 | 1.2/1.2/1.2 | 0.8/1.0/1.4 |

**Figure 12** shows CT images of the chest phantom acquired with simple PCCT and 5bin spectroscopic PCCT, using Fe-DBA and Ho-DBA. The signal non-uniformities and beam hardening artifacts are apparent in simple PCCT with Fe-DBA.

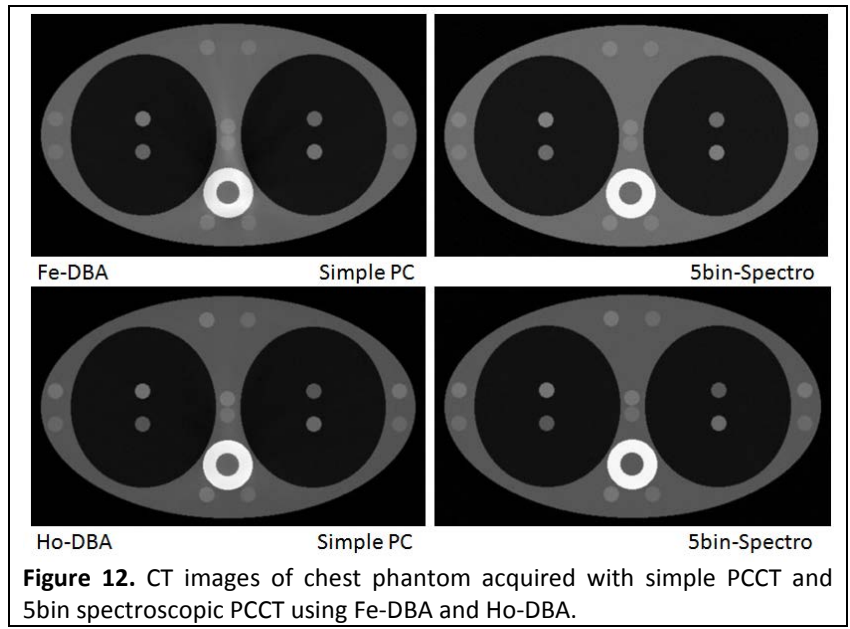

**Figure 12.** CT images of chest phantom acquired with simple PCCT and 5bin spectroscopic PCCT using Fe-DBA and Ho-DBA.

The signal non-uniformities and artifacts are better visualized in the image profiles (**figure 13**). The advantages of Ho-DBA versus Fe-DBA, and of 5bin spectroscopic PCCT versus simple PCCT are clearly observed.

**Table 5** includes the CT numbers (HU) of the contrast elements along with relative errors. The relative errors of the CT numbers of iodine and CaCO$_3$ are 32% and 44%, respectively, for simple PCCT with Fe-DBA. The errors are decreased to 2.4% and 3.9%, respectively, for 5bin spectroscopic PCCT. The simple PCCT with Ho-DBA provides 9.1% and 16% relative errors for iodine and CaCO$_3$, respectively. The 5bin spectroscopic PCCT provides only 0.7% and 1.1% relative errors for iodine and CaCO$_3$, respectively.

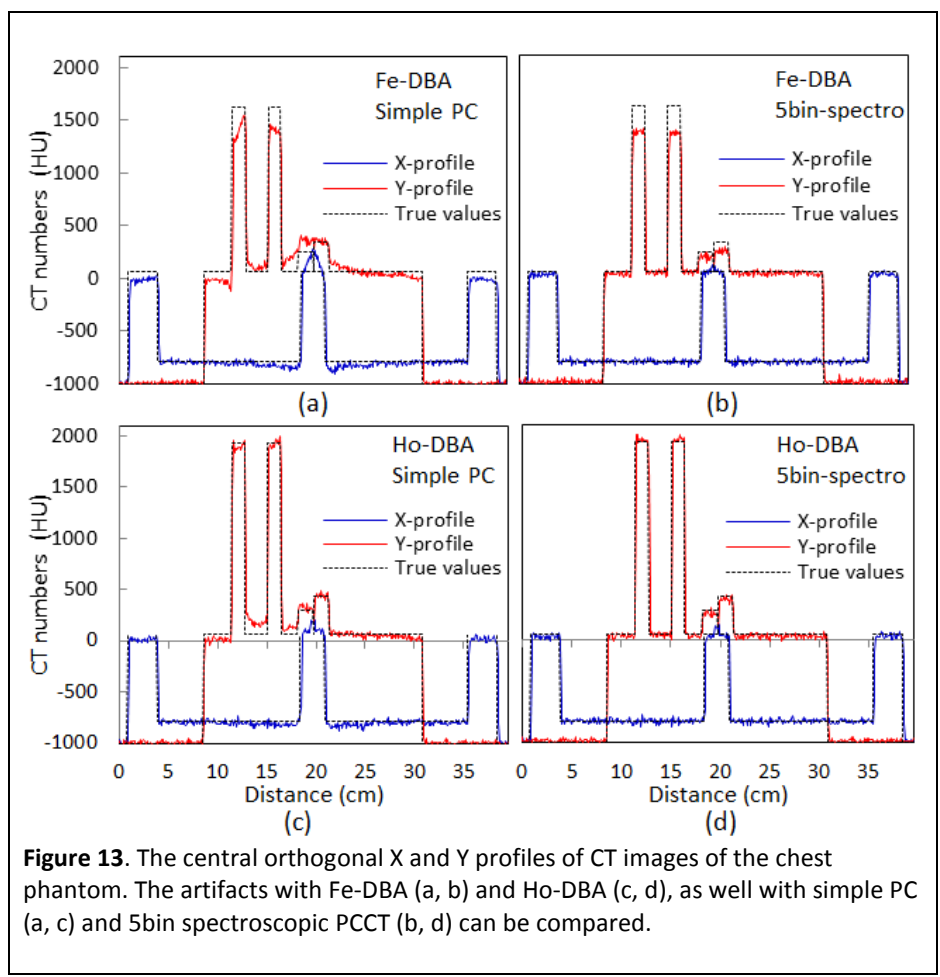

**Figure 13**. The central orthogonal X and Y profiles of CT images of the chest phantom. The artifacts with Fe-DBA (a, b) and Ho-DBA (c, d), as well with simple PC (a, c) and 5bin spectroscopic PCCT (b, d) can be compared.

For soft tissue background, the relative errors of CT numbers with simple PCCT, and Fe-DBA and Ho-DBA are as large as 140% and 65%, respectively. However, when 5bin spectroscopic PCCT is used, these errors are decreased to 10% and 4% with Fe-DBA and Ho-DBA, respectively.

**Table 5.** CT numbers and CT number errors in PC chest CT with DBA

| CT number (HU) | Fe-DBA Simple PC | Fe-DBA 5bin-Spectro | Ho-DBA Simple PC | Ho-DBA 5bin-Spectro | no DBA Simple PC | no DBA 5bin-Spectro |
|---|---|---|---|---|---|---|
| <I> | 244 | 255 | 398 | 413 | 387 | 359 |
| ΔI | 77 | 6 | 36 | 3 | 13 | 2 |
| **ΔI /<I>** | **0.316** | **0.024** | **0.091** | **0.007** | **0.034** | **0.006** |
| <Ca> | 212 | 203 | 269 | 273 | 265 | 250 |
| ΔCa | 94 | 8 | 43 | 3 | 12 | 1 |
| **ΔCa /<Ca>** | **0.443** | **0.039** | **0.160** | **0.011** | **0.045** | **0.004** |
| <ST> | 40 | 49 | 40 | 50 | 56 | 52 |
| ΔST | 56 | 5 | 26 | 2 | 7 | 1 |
| **ΔST/<ST>** | **1.400** | **0.102** | **0.65** | **0.040** | **0.125** | **0.019** |

The material decomposed (background cancelled) images of the chest phantom acquired with 5bin spectroscopic PCCT are shown in **figure 14**. Substantial deviations of background from zero are observed for Fe-DBA. The CNR and relative errors of the contrast are included in the **Table 6**. The Fe-DBA and Ho-DBA provide average CNR of 4.8 and 13.4, respectively, for iodine, and CNR errors are 9.4% and 1.6%.

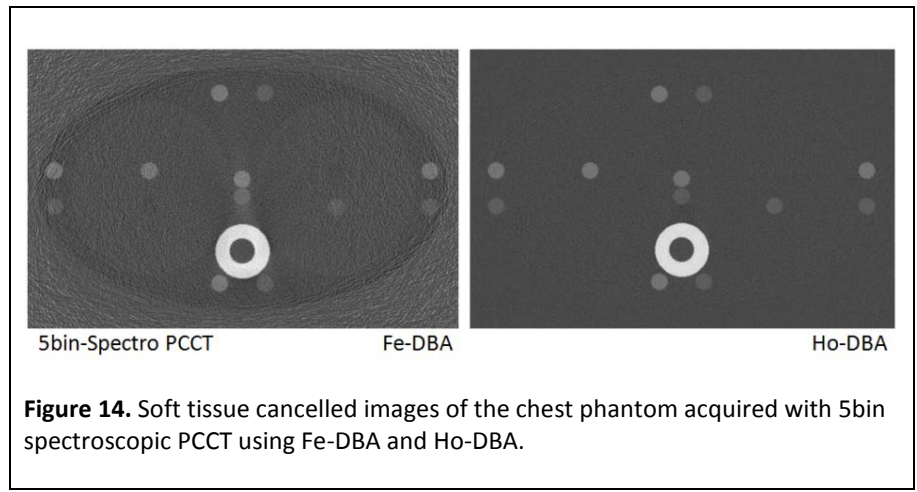

**Figure 14.** Soft tissue cancelled images of the chest phantom acquired with 5bin spectroscopic PCCT using Fe-DBA and Ho-DBA.

**Table 6.** CNR and relative contrast errors in tissue cancelled images

| CNR C -contrast | Fe-DBA Iod | Fe-DBA CaCO$_3$ | Ho-DBA Iod | Ho-DBA CaCO$_3$ | no DBA Iod | no DBA CaCO$_3$ |
|---|---|---|---|---|---|---|
| <CNR> | 4.8 | 2.2 | 13.4 | 5.6 | 10.2 | 4.9 |
| ΔC/<C> | 0.094 | 0.150 | 0.016 | 0.026 | 0.012 | 0.021 |

**Figure 15** shows that as the low energy threshold is increased the artifacts are decreased because more hardening-sensitive part of the spectrum is left below threshold. Interestingly, with Ho-DBA, 58keV threshold provides larger artifacts than 40keV threshold. This is due to Ho K-edge: the low energy photons are better preserved with 40keV threshold than with 58keV. Also, the beam hardening artifacts with energy integrating CT is smaller than with simple PCCT. This is due to energy weighting effect when PCCT weights low energy photons higher than the energy integrating CT (Shikhaliev 2005).

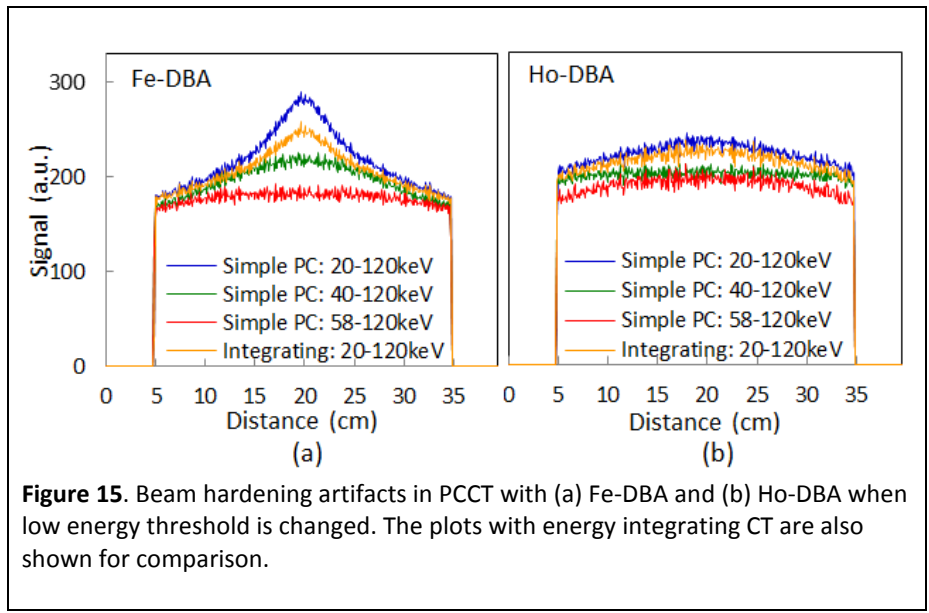

**Figure 15**. Beam hardening artifacts in PCCT with (a) Fe-DBA and (b) Ho-DBA when low energy threshold is changed. The plots with energy integrating CT are also shown for comparison.

## 4. Discussion and conclusion

Spectroscopic PCCT has a potential for minimizing image noise by optimal photon energy weighting and electronics noise rejection (Shikhaliev 2008). Using appropriate DBA can further increase CNR by minimizing detected scatter and better separating low and high energy bins (Shikhaliev 2012; Atak and Shikhaliev 2016), which would enable material selective CT at lowest patient dose (Atak and Shikhaliev 2015). The PCCT with DBA was investigated in this work and following are the findings: **(1)**Simple PCCT with DBA exhibits major beam hardening artifacts and quantification errors; **(2)**2bin spectroscopic PCCT with DBA provides some improvement, but residual artifacts and errors are still large; **(3)**5bin spectroscopic PCCT with DBA exhibits minimal artifacts and errors associated with DBA; **(4)**DBA made of K-edge material (Ho in this case) performs nearly as good as DBA based on soft tissue equivalent material, and both DBA perform much better than Fe-DBA; **(5)**Although beam hardening artifacts are minimal when Fe-DBA is used in 5bin spectroscopic PCCT, the photon starvation noise is a problem because Fe-DBA strongly absorbs photons in low energy bins; **(6)**K-edge DBA (Ho-DBA in this case) can be much smaller (10 times smaller for Ho) than Fe-DBA and better practical implementation; **(7)**Stepwise DBA exhibits "edge" artifacts in addition to "roof" artifacts observed with smooth DBA; **(8)**All types of artifacts are nearly eliminated when Ho-DBA is used in 5bin spectroscopic CT, and no photon starvation noise is observed; **(9)**Conventional energy integrating CT with DBA also exhibits major "roof" artifacts and quantification errors, although slightly less than in simple PCCT with DBA.

The quantitative data were included in: **Table 1** – Simulated parameters of the DBA; **Table 2** – Magnitudes of beam hardening artifacts with DBA; **Table 3** – Noise non-uniformities with DBA; **Table 4** – CNR and CNR non-uniformities in decomposed images; **Table 5** – CT numbers and CT number errors; and **Table 6** – CNR and contrast non-uniformities in decomposed images.

As noticed above, beam intensity modulation was previously investigated for projection x-ray imaging (Hasegawa et al. 1986; Vlasbloem and Kool 1988; Xu et al. 2004). However, it did not find routine clinical applications due to the mechanical complexity and instability that may also exist for PCCT with DBA. Modern CT systems operate at approximately 360 degree rotation in 0.5s. In our simulations DBA changed its shape smooth (720 ties/360 degree) and stepwise (32 times/360 degree), and no measurable differences were found in artifact patterns and magnitudes. Other works used 15 times reshaping of DBA in 360 degree (Hsieh and Pelc 2013; Szczykutowicz and Mistretta 2013; Szczykutowicz and Mistretta 2013; Hsieh et al. 2014; Szczykutowicz and Mistretta 2014). Assuming 15 times reshaping per gantry rotation, the DBA must reshape 30 times per second. This may be a challenging task although DBA fabricated from K-edge material could have smaller sizes and simplify this task. Notice also that many K-edge materials are readily available and can be machined and shaped (Shikhaliev 2012; Shikhaliev 2012; Atak and Shikhaliev 2016).

Another potential challenge is accurate calibration of the DBA motion. The CT image acquisition with DBA includes three steps: (1)Low-dose pre-scanning of the patient without DBA to calculate needed DBA profiles; (2)Scanning patient with DBA on the beam; and (3)"Air scanning" without patient and with DBA on the beam. The $1^{st}$ and $2^{nd}$ scans with patient on the beam do not have to be in a single breath hold as they are relatively immune to slight motions of the patient (in the order of centimeter). This is because DBA steps are relatively large, and information about DBA profile is cancelled out during the image reconstruction. However, the $3^{rd}$ scan should repeat the dynamic reshaping patterns of the DBA from the $2^{nd}$ scan with high accuracy. Any discrepancy between these two patterns may result in artifacts. Therefore, high accuracy, precision, and stability of the DBA motion are necessary.

In the current work we focused on general imaging and material selective imaging performances of PCCT with DBA. We did not investigate dose optimization and scatter reduction potentials with DBA as these were studied in previous works (Hsieh and Pelc 2013; Szczykutowicz and Mistretta 2013; Szczykutowicz and Mistretta 2013; Hsieh et al. 2014; Szczykutowicz and Mistretta 2014; Hsieh and Pelc 2015). We further assumed an ideal PC and energy integrating detectors that in practice may have a number of limitations. The PC detectors suffer from suboptimal detection efficiency, pixel crosstalk due to charge sharing, carrier trapping, and suboptimal energy resolution due to statistical fluctuation of created charge (d'Aillon et al. 2006; Kuvvetli and Jørgensen 2007; Shikhaliev et al. 2009; Faby et al. 2016). Currently, no detector model exists that could combine the above limitations of the PC detectors and correctly predict the detector response. The energy integrating detectors also exhibit similar limitations (Heismann et al. 2008). Therefore, using ideal detector model in this work was justified because it was necessary to single out and quantify the basic positive and negative effects associated with using DBA.

Based on the patient's attenuation map, the DBA generates x-ray intensity profiles such that the count rates of the detector pixels are not higher than certain threshold. In the current study, we set this threshold to be equal to the lowest count rate among all pixels at all projection angles during one full gantry rotation. This minimal count rate occurred when the ray passed through most attenuating part of the object and was 2.2Mc/pixel/sec for cylindrical CT phantom and 4.8Mc/pixel/sec for the chest phantom. The count rate threshold could be set higher than the above minimal values provided that the PC detector can handle such high count rates (Hsieh and Pelc 2015). Although the used count rate thresholds resulted highest beam attenuations by DBA (see **Table 1**), it is still acceptable for most imaging applications (Atak and Shikhaliev 2016).

The DBA provides spatially variable HVL of the x-ray beam, and the average HVL for Fe-DBA, Ho-DBA, and ST-DBA were 10.3mm, 7.1mm, and 6.8mm Al equivalent, respectively. These HVL are comparable to the HVL used in current CT systems (NHS 2009), with the exception that HVL with Fe-DBA was slightly higher than maximum HVL of 9.1mm used in current CT systems (NHS 2009). To minimize unnecessary x-ray attenuation, the 120kVp x-ray beam was minimally filtered before the DBA and had 4.5mm Al equivalent HVL. Therefore, in areas where DBA thickness was zero, the HVL at the phantom surface was 4.5mm Al equivalent. This 4.5mm minimal HVL was chosen to meet the FDA requirements for minimal HVL which is 4.3mm Al equivalent for 120kVp beams (FDA-DHHS 2014).

It is concluded that although Fe-DBA may introduce substantial image artifacts and quantification errors, the DBA fabricated from K-edge material (such as Ho used in this study) can minimize these artifacts and errors, improve material decomposition due to better separation of the low and high energy bins, and provide miniature sizes. However, engineering challenges may still exist with fabrication and applications of DBA, and further investigations are needed.

## 5. Acknowledgements